# LA PRECESIÓN DE LOS EQUINOCCIOS

## Una propuesta pedagógica para docentes de la escuela secundaria

**Alejandro Gangui**
*Instituto de Astronomía y Física del Espacio, UBA-CONICET*

*Sumario para* **Contenidos**

Segunda parte de una secuencia didáctica sobre temas de astronomía destinada a docentes de la escuela secundaria.

Esta nota es la segunda y última parte de la secuencia didáctica sobre temas de astronomía cuya primera parte fuera presentada en el número anterior ('El movimiento de los cielos', CIENCIA HOY 106, agosto-septiembre 2008, pp.58-62). Se recomienda que el docente se familiarice con el primer texto antes de abordar el segundo.

Igual que la anterior, esta secuencia didáctica sobre algunas cuestiones de astronomía está destinada a docentes de la escuela secundaria, para que la lleven al aula. A diferencia de la anterior, esta nueva propuesta apunta a docentes de los últimos años. Su tema central, a ser presentado en forma progresiva, es el *movimiento de precesión de los equinoccios*. Entre las actividades que se sugieren, se cuenta que los alumnos construyan un modelo o maqueta simple de un mini sistema solar y una banda zodiacal de estrellas que lo rodee. Esa construcción, junto a la discusión entre pares y con el docente, los ayudará a familiarizarse con aspectos astronómicos a veces tratados muy por encima en la escuela, o sencillamente omitidos, por falta de recursos de fácil empleo.

La primera secuencia incluyó una serie de actividades simples orientadas a comprender el ciclo día-noche, el movimiento anual de la Tierra alrededor del Sol, la inclinación del eje de rotación de nuestro planeta y la causa de las estaciones del año. Asimismo, sugirió una construcción sencilla de un sistema solar simplificado, el que también será empleado aquí. Las tres actividades delineadas en lo que sigue poseen un grado de complejidad un poco mayor, pero son la continuación natural de las presentadas antes, por lo que no es aconsejable encarar estas sin haber realizado aquellas. Unas y otras, pero sobre todo estas, apuntan a alumnos de la secundaria con alguna predisposición, y quizás también cierto entrenamiento, para el pensamiento abstracto y reflexivo, y para la construcción de modelos.

Tomaremos en cuenta aquí la forma real de la Tierra, cuya rotación la apartó de una esfera perfecta, y consideraremos el sutil bamboleo de su eje de rotación, el cual hace que todas las estrellas del cielo nos parezcan irse desplazando año tras año de sus posiciones. Entre las estrellas se cuentan, por supuesto, las que indican los puntos equinocciales, es decir los lugares del cielo que, vistos desde la Tierra, coinciden con la posición del Sol en el comienzo de la primavera y del otoño, o, como es más frecuente decirlo, los lugares por los que 'pasa' el Sol en esos momentos. Este es el origen de la expresión precesión de los equinoccios.

El autor espera que los alumnos –guiados por los docentes– puedan abordar estos temas algo abstractos de astronomía partiendo de sus propios conocimientos y profundizando los adquiridos para contrastarlos con los anteriores. La maqueta a construir en la última actividad permitirá entender cómo los astros –objeto de muchos mitos y tradiciones– dieron origen a los signos del zodíaco, creados contemplando la bóveda estrellada del cielo. Los alumnos podrán también entender cómo las constelaciones que se ven en cada estación del año (o, mejor dicho, las estrellas que las forman) se van desplazando en forma lenta pero continua. Ello fue así a lo largo de la extensa historia pasada de la humanidad, y así será en el futuro.

Como lo expresó en la nota anterior, el autor espera que los docentes curiosos e interesados por la astronomía puedan, por lo menos, aprovechar una parte de sus propuestas, independientemente del año en que enseñen, del nivel socio-económico de la escuela y de los restantes temas tratados en esta u otras áreas de enseñanza. Mientras se observa, se discute, se reflexiona y se lleva a cabo una construcción simple, la secuencia de actividades sugerida no es más que una excusa para hablar sobre astronomía.

## Material de trabajo

Para llevar adelante esta propuesta se requiere disponer del siguiente material:

* La maqueta del mini sistema solar construida con motivo de la propuesta pedagógica anterior (las indicaciones precisas para construirla están en la nota precedente).
* Una esfera de telgopor, de unos 12cm de diámetro aproximadamente.
* Una varilla delgada de madera, de sección circular, de unos 16cm de largo, con sus extremos afilados para poder clavarla en el telgopor.
* Un trompo y algunas bombitas de agua o globos pequeños.

## Secuencia de actividades

### Actividad 1: La forma real de la Tierra

Esta actividad tiene como objetivo reflexionar sobre la verdadera forma de la Tierra. Se busca que los alumnos reconozcan que determinadas variables, como la velocidad de rotación de un astro, pueden modificar aspectos significativos de este.

Para comprender la forma real de la Tierra, el docente podrá proceder por analogía, recurriendo a una bombita de agua (de las populares en Carnaval) e indicando a los alumnos que imaginen que ella representa a la Tierra. Si su tamaño no es demasiado grande, la bombita conservará una forma aproximadamente esférica, pero si la hacemos rotar rápidamente sobre la mesa, veremos sin dificultad que se deforma, pues se achata en los polos (por donde pasaría su eje de rotación) y se hace más protuberante en sus regiones ecuatoriales.

Se recomienda al docente fomentar en los alumnos la expresión de las ideas con que llegaron a clase, y alentar la discusión de ellas en grupo. De revelarse algún conflicto entre esas ideas y lo que los mismos estudiantes deducen en la discusión y reflexión grupales, el docente podrá hacer converger sus concepciones hacia los modelos y conceptos que hoy emplea la ciencia. También podrá sugerir reflexionar sobre la forma real de la Tierra representándola por la esfera de telgopor de 12cm (atravesada por la varilla de 16cm, que indicaría su eje de rotación). De disponerse de la maqueta del mini sistema solar sugerida para la primera secuencia, podrá resultar también útil. Se puede colorear la esfera de la Tierra con un tono cualquiera (verde, por ejemplo) y oscurecer una banda ecuatorial, que representaría el abultamiento real de la Tierra.

Con los alumnos divididos en grupos pequeños, el docente puede plantear algunas preguntas para guiar la discusión y, especialmente, alentarlos a que argumenten entre ellos. Por ejemplo:

- * Hagamos girar la bombita velozmente sobre la mesa. ¿Qué observan en su forma? ¿Cómo lo explican? Justifiquen sus respuestas.
- * ¿Es la Tierra un cuerpo completamente rígido? ¿O es que su propia estructura interna, y sus océanos y atmósfera, la convierten en un objeto deformable?
- * ¿Cómo creen que es la forma de la Tierra? ¿Piensan que es realmente una esfera?
- * Sabemos que la Tierra, al girar sobre su eje, da una revolución completa cada 24 horas (y podemos calcular que, en la superficie terrestre, en el Ecuador, la velocidad de rotación es de unos 1000 km/h). En esta situación, ¿la Tierra seguirá siendo una esfera? ¿Por qué?

Luego de lanzadas las preguntas anteriores, de escuchar a los alumnos y de pedirles que registren sus respuestas individuales en sus cuadernos (para poder confrontarlas con las ideas a que arribe el grupo), el docente puede leer a toda la clase el texto de Voltaire que reproducimos, extraído de su decimocuarta Carta filosófica, en el que el autor menciona las dos formas de concebir el mundo en su época. Los ingleses, siguiendo a Newton, se inclinaban por imaginarse un universo vacío, en el que los cuerpos celestes estaban sujetos a unas fuerzas gravitatorias cuyos orígenes no resultaban claros. Los franceses, siguiendo a Descartes, concebían un universo lleno de una materia sutil y desconocida cuyas corrientes y torbellinos arrastraban a los cuerpos celestes. Entre las divergencias de concepción que señala Voltaire, también está la forma de la Tierra.

Sabemos hoy que el modelo del inglés terminó predominando, y que la Tierra, considerada en tiempos geológicos, puede concebirse como una pelota de material sólido deformable y viscoso que rota rápidamente sobre su eje. Está, efectivamente, algo achatada en los polos y, por lo tanto, no tiene la forma de un melón. Pero el achatamiento polar es leve, pues solo alcanza a un 0,34%: la circunferencia polar es apenas unos 21,7km más corta que la ecuatorial. Ese achatamiento es tan tenue que no resulta discernible desde el espacio, a diferencia de lo que sucede con Saturno o Júpiter, cuyos respectivos achatamientos se advierten incluso con un telescopio pequeño. La constitución gaseosa de las partes más externas de esos dos planetas gigantes ayuda mucho a que su deformación sea más pronunciada.

*Recuadro para la actividad 1*

### Voltaire y la forma de la Tierra

Un francés que llega a Londres encuentra las cosas muy cambiadas en filosofía, como en todo lo demás. Ha dejado el mundo lleno; se lo encuentra vacío. En París, se ve un universo compuesto de torbellinos de materia sutil; en Londres, no se ve nada de eso. Entre nosotros, es la presión de la Luna la que causa el flujo del mar; entre los ingleses, es el mar el que gravita hacia la Luna […]. Notaréis además que el Sol, que en Francia no interviene para nada en este asunto, contribuye aquí por lo menos en una cuarta parte. Entre vosotros, cartesianos, todo sucede a través de una presión de la que nada se comprende; para monsieur Newton, es por una atracción cuya causa no se conoce mejor. En París, os figuráis la Tierra hecha como un melón; en Londres, está aplastada por los dos lados (Voltaire, *Cartas filosóficas*, Alianza, Madrid, 1988).

### Actividad 2: La atracción de la Luna sobre la Tierra y el movimiento del eje terrestre.

En esta actividad se discute la fuerza de atracción gravitatoria que sufren todos los cuerpos masivos y cómo la interacción entre la Tierra y la Luna (y, en menor medida, entre aquella y los demás cuerpos astronómicos) altera la orientación en el espacio del eje de rotación de la Tierra. Se busca que los alumnos reflexionen sobre diversos aspectos de la gravitación en una aplicación concreta que, aunque poco intuitiva, tiene gran importancia en astronomía.

El concepto importante a transmitir a los estudiantes es la perturbación causada por la atracción gravitatoria lunar en la orientación en el espacio del eje de rotación de la Tierra. Este, que como vimos en la secuencia anterior, no es vertical con respecto plano de la órbita terrestre, sufre una alteración adicional por la existencia del abultamiento ecuatorial de nuestro planeta, discutida en la actividad 1.

Para transmitir estos conceptos, en primera instancia el docente podrá mostrar a toda la clase la mayor fuerza de atracción de la Luna sobre la región ecuatorial de la Tierra más próxima a aquella, comparada con la menor fuerza gravitatoria ejercida por el satélite terrestre sobre la región ecuatorial de la Tierra más alejada de la Luna. Para ello conviene hacer uso del pizarrón e ilustrar las analogías que existen entre la Tierra y un trompo en rotación, que se representan en la figura 1. De disponer de algunos trompos, el docente podrá hacer que los alumnos los hagan girar en sus mesas o en el piso: verán así cómo es su movimiento completo, que incluye tanto el rápido movimiento de rotación sobre su eje como el movimiento de bamboleo del propio eje.

El trompo, que inicialmente gira con su eje perpendicular al plano del piso, gradualmente pierde velocidad por rozamiento con el aire y con el piso en el punto de contacto. Cualquier leve perturbación lo hace inclinarse para un costado, con lo que comienza su caída. Sin embargo, la inercia de su movimiento giratorio (técnicamente llamada *conservación del momento angular*), trata de mantener la posición del eje de rotación ante cualquier fuerza que la pueda alterar, como su peso. Por ello el trompo no cae sin más, sino que su eje describe un movimiento circular de oscilación o bamboleo. Los extremos del eje del trompo comienzan a describir una circunferencia en el espacio (y el eje describirá una suerte de doble cono). Ese movimiento del eje del trompo se denomina *precesión* y se dice que el eje *precesa*.

El caso de la Tierra es análogo, con la diferencia de que no hay rozamiento que frene su movimiento de rotación, pues nada hay con qué rozar en el espacio interplanetario. A eso debemos agregar la inmensa masa del planeta, que incrementa en forma notoria su inercia de giro. No es fácil modificar la velocidad de rotación de semejante masa planetaria, de la misma forma que no es simple alterar la orientación del eje terrestre. Se requiere otro cuerpo astronómico de gran porte, o muy cercano, para lograrlo. La Luna cumple esa función.

Por estas razones, el eje de la Tierra está sujeto un movimiento de precesión que altera su dirección en el espacio, como lo muestra la figura 2. En un trompo, ese movimiento circular del eje puede durar algunas decenas de segundos. Al eje de la Tierra le lleva aproximadamente 26.000 años hacer un giro completo. Los polos norte y sur celestes son los lugares del firmamento o de la bóveda de las estrellas a los que apuntan las prolongaciones del eje de rotación de la Tierra por encima de los polos terrestres. Esos polos celestes describen en ese lapso un movimiento de 360º con respecto al telón de fondo inmóvil de las estrellas (o, lo que es lo mismo, las estrellas describen ese movimiento con relación al observador terrestre). Por ello, *Polaris* no fue ni será siempre la estrella polar norte.

¿Y qué sucede con los equinoccios, incluidos en el concepto de precesión de los equinoccios? ¿Y con los solsticios, un concepto inseparable del de los equinoccios? Antes que nada, ¿qué son equinoccios y solsticios? El hecho de que el eje de rotación terrestre no sea perpendicular al plano definido por el movimiento de traslación de la Tierra alrededor del Sol (o plano de la eclíptica) tiene como consecuencia que el plano del ecuador terrestre (que es perpendicular al eje de rotación de la Tierra y pasa por el centro de esta) no coincida con el de la eclíptica. Por la misma razón, el plano ecuatorial celeste (o sea, la extensión al cielo del plano del ecuador terrestre, hasta su intersección con la bóveda de las estrellas) no es paralelo al plano de la eclíptica. La intersección de ambos planos es una recta que pasa por el centro de la Tierra y alcanza la bóveda celeste (figura 3). Los dos puntos de intersección de esa recta con dicha bóveda se denominan puntos equinocciales o equinoccios (y están tanto sobre el ecuador celeste como sobre la eclíptica, a 180º uno del otro). Si desde los equinoccios nos desplazamos por la esfera celeste a lo largo de la eclíptica, podemos definir dos puntos equidistantes de cada uno de aquellos (es decir, a 90º de cada equinoccio), uno por encima del ecuador celeste y el otro por debajo. Esos son los dos solsticios, uno del verano boreal (e invierno austral) y el otro del verano austral (e invierno boreal). Cuando el Sol, en su movimiento anual aparente, coincide con alguno de estos cuatro puntos de la bóveda celeste, nuestro calendario marca un solsticio o un equinoccio, y comienzan en la Tierra las correspondientes estaciones del año.

De la misma manera que los polos norte y sur celestes se desplazan con respecto a las estrellas lejanas, también los solsticios y equinoccios, esos cuatro puntos notables de la eclíptica, cambian, muy lentamente, su posición en la bóveda celeste o posición zodiacal.

Sabemos que la atracción gravitatoria entre dos cuerpos, o entre zonas de esos cuerpos, es tanto más grande cuanto menor es la distancia que los separa. Sabemos también que la órbita de la Tierra alrededor del Sol está contenida en un plano, llamado el plano de la eclíptica, y que la Luna describe una órbita, también plana, alrededor de la Tierra. Estos dos planos se hallan muy próximos entre sí: con centro común en la Tierra, se inclinan uno respecto al otro apenas unos 5º. Por ello, y para simplificar la discusión, podemos considerar que la Luna y el Sol interactúan con la Tierra desde un único y mismo plano, que tomaremos como el plano de la eclíptica.

Las siguientes sugerencias y preguntas pueden resultar útiles al docente para guiar la discusión:

* Dado que el eje de rotación de la Tierra se halla inclinado con respecto al plano de la eclíptica, dibujen cualitativamente las fuerzas sobre cada abultamiento ecuatorial de la Tierra: sobre el que se enfrenta a la Luna y sobre el que se halla opuesto a ella. (Hacer los dibujos en dos dimensiones, sobre un papel, por ejemplo. Ver figura 1).
* ¿Sobre qué parte de la Tierra es mayor la fuerza gravitatoria de la Luna? ¿Sobre qué parte de nuestro planeta es comparativamente menor?
* Esas fuerzas, aplicadas sobre distintas partes de la Tierra, producen un *momento* o *torque* (una fuerza de rotación, análoga a la que producimos con las manos cuando abrimos una canilla: hacemos que la válvula de esta gire pero no desplazamos a la canilla de su lugar). Dicho torque, ¿tiende a inclinar el eje terrestre o a enderezarlo?
* ¿Encuentran alguna similitud entre lo que sucede con un trompo en rotación y lo que sucede con la Tierra en rotación?
* ¿Hacia dónde cae el trompo? ¿Qué es lo que lo hace caer?
* ¿Hacia dónde 'cae' la Tierra? ¿O es que tiende a girar? ¿Qué cuerpo astronómico produce esta 'caída'? Justifiquen sus respuestas. (Véase el recuadro: 'La Luna, más que el Sol, modifica la orientación del eje terrestre'.)
* Discutan: ¿qué significa 'caer'? ¿Por qué un satélite no se cae? ¿Y la Luna? ¿No se cae, o está constantemente cayendo y nunca termina de caer?
* Por último, imaginemos ahora que la Tierra no posee ese abultamiento ecuatorial originado por su rápida rotación. Imaginando la Tierra como una esfera perfecta, ¿podríamos dibujar pares de fuerzas de interacción con la Luna que la hagan inclinarse para algún lado? ¿O es que para cada fuerza que podamos imaginar existe otra que anula exactamente su efecto? (Sabemos que para una Tierra esférica, el movimiento de precesión no existe.)
* Discutan: ¿hay alguna manera de inclinar o rotar a una esfera perfecta? Si es perfecta, ¿cómo podemos saber que rota, o que no rota? ¿Tiene sentido hablar de la rotación de una esfera perfecta?

*Recuadro para la actividad 2*

> **La Luna, más que el Sol, modifica la orientación del eje terrestre**
>
> A la Luna le cabe la principal responsabilidad en el cambio de orientación en el espacio del eje de rotación terrestre. Los demás cuerpos astronómicos, incluyendo el Sol, tienen una influencia menor. Aproximadamente, dos tercios del efecto corresponden a la acción de Luna y un tercio al Sol. Sin embargo, el Sol ejerce sobre la Tierra una acción gravitatoria mucho mayor que la Luna. Después de todo, la Tierra se traslada alrededor del Sol, por efecto de la atracción gravitatoria de este, y no alrededor de la Luna. ¿Cómo se explica esta aparente contradicción?
>
> La acción gravitatoria de cada uno de estos astros sobre la zona ecuatorial de la Tierra se puede concebir como una *fuerza diferencial neta*, es decir, como la diferencia entre las acciones gravitatorias que ejercen dichos astros en dos puntos de la superficie de la Tierra ubicados sobre el ecuador y en las antípodas uno con respecto al otro (en posiciones opuestas con respecto al centro de la Tierra). Esos puntos distan entre sí unos 10.000km, muy poco en comparación con los 400.000km que nos separan de la Luna o los 150 millones de kilómetros que dista la Tierra del Sol.
>
> La fuerza diferencial neta que actúa en esta situación no disminuye con la inversa del cuadrado de la distancia, como lo establece la fuerza clásica de Newton, sino –aproximadamente– con el cubo de la distancia entre la Tierra y el respetivo astro. El motivo de esto último radica en que el término newtoniano clásico (el cuadrático) es casi idéntico si se lo calcula entre un punto de la superficie terrestre y el astro, o entre el punto antípoda de la Tierra y el mismo astro (como dijimos, 10.000km es una distancia muy pequeña comparada con, por ejemplo, 400.000km).
>
> Estos dos términos cuadráticos de la interacción son aproximadamente idénticos en magnitud, pero tienen sentidos contrarios en sus efectos sobre la Tierra (uno trata de enderezarla; el otro la tiende a inclinar aun más). Es por ello que se compensan y se vuelve necesario calcular órdenes superiores en la interacción (de otra manera, el resultado sería muy burdo y daría cero). El término siguiente es el cúbico y, por ello, la interacción diferencial decae muy rápidamente con la distancia entre los cuerpos. Así, la gran proximidad de la Luna, comparada con la del Sol, compensa con creces su débil masa (también comparada con la del Sol). En consecuencia, igual que sucede con las mareas, la interacción de la Luna es el principal responsable del cambio de orientación del eje terrestre que da origen al movimiento de precesión de los equinoccios.

**Actividad 3: El movimiento de precesión de los equinoccios y el corrimiento de las constelaciones del zodíaco.**

En esta actividad se ponen en juego todos los conocimientos adquiridos en las etapas precedentes. Consiste en la construcción de una maqueta en la que estén representadas las estrellas de la banda zodiacal que rodean al sistema solar simplificado de la maqueta descripta en la anterior nota didáctica, publicada en el número 106 de CIENCIA HOY. Se busca que los alumnos reflexionen sobre el movimiento de precesión y sobre sus consecuencias observables, y que planteen hipótesis sobre qué aspecto tendrá el cielo nocturno en distintas épocas –pasadas y futuras– de la humanidad. Por último, se incluyen algunos elementos útiles para discutir en clase ciertos aspectos de la astrología.

Conviene comenzar esta actividad repasando conocimientos sobre los que versó la anterior nota didáctica. Con ayuda de la maqueta allí descripta e invitando a que los alumnos desplacen la Tierra en órbita alrededor del Sol, los docentes pueden ayudar a reflexionar sobre las distintas regiones de la bóveda celeste visibles de noche en las cuatro estaciones del año. Recuérdese que (prescindiendo de la precesión) la orientación del eje terrestre permanece constante, con sus extremos apuntando siempre en las mismas direcciones del espacio: el norte hacia la estrella *Polaris* y el sur hacia una estrella llamada *Sigma Octantis*, que puede identificarse en un mapa del cielo austral, en caso de tenerlo. El brillo aparente de la segunda estrella es bastante inferior al de Polaris y, hoy, se halla un poco más alejada del polo sur celeste de lo que está Polaris del norte (esta tampoco está *exactamente* en el polo norte celeste).

Divididos en grupos pequeños, el docente podrá guiar a sus alumnos para que también recapitulen el significado de los acontecimientos astronómicos singulares generados por la órbita terrestre, como solsticios y equinoccios, inicio de las cuatro estaciones del año, asociados con la posición de la Tierra en su órbita y con la inclinación de su eje de rotación.

El segundo paso es que los alumnos, guiados por el docente y ayudados por la maqueta, imaginen la bóveda celeste que rodea su maqueta. El propósito es ayudarlos a entender que el fondo de las estrellas visibles cambia a medida que avanza el año. Las estrellas que se ven de noche desde la Tierra están en dirección aproximadamente contraria al Sol, pues la cara terrestre con iluminación diurna apunta al Sol y la otra mira en sentido contrario. Las estrellas ubicadas en la dirección del Sol, es decir, aquellas hacia las que apunta la cara iluminada de la Tierra, no se pueden distinguir, como es obvio, porque el Sol no permite que se vean. Esto sucede porque su luminosidad es débil comparada con la del Sol, y porque la atmósfera dispersa la luz solar. Venus (también llamado el lucero, pero que no es una estrella fija sino un planeta que se desplaza y no emite luz propia sino que refleja la del Sol) a veces llega a verse de día, porque está entre los objetos más luminosos. En la Luna, por su lado, que no tiene una atmósfera como la de la Tierra, se ve el cielo tapizado de estrellas incluso de día (es decir, desde la cara iluminada del satélite).

Por estos caminos los alumnos podrán llegar a entender el concepto de *posición aparente del Sol en la bóveda celeste*, cuyos cambios se deben a la trayectoria anual de traslación de la Tierra en órbita alrededor del Sol, algo completamente diferente del movimiento diario del Sol en el cielo por efecto de la rotación de la Tierra. También podrán entender qué es el *zodíaco*, es decir, la sucesión de las constelaciones que se verían en el cielo diurno detrás del Sol (si los rayos de este no lo impidieran); y qué significa la expresión astrológica de que el Sol 'reside' en determinada constelación en tal época del año. También en la comprensión de estos conceptos está la clave para apreciar por qué la fecha de nuestro cumpleaños es la menos adecuada para ver en el cielo las constelaciones asociadas con nuestro signo astrológico de nacimiento (como escribimos en 'Dante, astronomía y astrología', CIENCIA HOY, 104: 30).

Las siguientes preguntas pueden ayudar a estimular la reflexión (la figura 4 ilustra la situación a que se refieren):

* Imaginen que pasan unos 13.000 años (la mitad de 26.000), ¿Cuál será la orientación del eje terrestre en esa época futura?
* ¿Qué cambios observaríamos entonces en el cielo? Expliquen sus respuestas.
* ¿Qué objetos constituyen puntos de referencia en el espacio?

Llegados a este punto de la actividad, es oportuno que el docente proponga completar la maqueta simplificada que los alumnos vinieron usando hasta ahora con una banda circular de estrellas, como lo indica la figura 5. Dicha banda representa una porción de la esfera celeste: aquella que contiene a la eclíptica y, por ende, a la trayectoria o camino anual aparente del Sol en el firmamento (es decir, en el cielo que vemos desde la Tierra). Como figura 6 hemos incluido tres tiras que, fotocopiadas y pegadas, forman la cinta cerrada a colocar rodeando el sistema solar simplificado. La banda incluye todas las estrellas prominentes de esa porción del cielo.

Con el dispositivo completo, el docente podrá discutir cómo es realmente el movimiento de los cielos. También podrá aclarar, si lo cree conveniente, el significado astronómico de los signos astrológicos del zodíaco, aunque con ello corre el peligro de que la atención de la clase se desvíe hacia un tema (el valor de la astrología) que es preferible evitar para que el esfuerzo de comprensión astronómica no se diluya.

Si decide desoír esta advertencia y aventurarse en terreno tan espinoso, tal vez pueda encontrar ayuda en el recuadro 'Astronomía y astrología' de nuestro artículo sobre Dante (CIENCIA HOY, 104: 30). También podría proponer –hacia el final de la clase– que los alumnos busquen información sobre esos temas y, en una clase posterior, organizar un debate entre defensores y detractores de las creencias astrológicas. A continuación algunas preguntas para discutir en grupo:

* ¿Cuáles son las constelaciones zodiacales visibles en el cielo nocturno alrededor de la fecha de nacimiento de cada uno?
* ¿Qué cambia en la respuesta anterior si viviéramos hace 13.000 años? ¿Y para alguien que nacerá dentro de 13.000 años? Justifiquen sus respuestas.
* Entre las constelaciones de Escorpio y Sagitario, la eclíptica cruza una constelación zodiacal que no pertenece al zodíaco tradicional. ¿Cómo se llama? ¿La eclíptica solo pasa por allí ahora o siempre lo hizo? ¿Depende este hecho de la época histórica? ¿Se modifica con la precesión?
* Contemos las constelaciones zodiacales. ¿Cuántas vemos en la banda de la maqueta?
* ¿En qué constelación 'reside' el Sol aproximadamente entre el 30 de noviembre y el 17 de diciembre?

Como comentario final, dejemos constancia de que la observación del cielo, característica de la astronomía, está mayormente ausente de las actividades que propusimos, por la índole de los temas abordados: la forma real de la Tierra, la atracción gravitatoria de los otros cuerpos celestes sobre ella y el muy lento movimiento de precesión del eje terrestre (y por consiguiente de solsticios y equinoccios). Este último movimiento es tan sutil e imperceptible en la escala de tiempos de una vida humana, que hizo falta un observador minucioso y perseverante, como Hiparco (siglo II aC), un cielo diáfano y puro, como el de su observatorio de Alejandría, y una saludable confianza en los documentos de trabajo de observadores anteriores, como los de Timocaris (320-260 aC) y su discípulo Aristilo, para no solo llegar a detectar las pequeñas anomalías en la posición de las estrellas, originadas por este efecto, sino también –y muy lentamente– comenzar a comprenderlas.

*Lecturas sugeridas*

CAMINO N y ROS R M, 1997, '¿Por dónde sale el Sol?', *Educación en ciencias*, 1, 3: 11-17.
GANGUI A, IGLESIAS M y QUINTEROS C, 2008, 'Diagnóstico situacional de los docentes de primaria en formación sobre algunos fenómenos astronómicos', *Memorias del I congreso internacional de Didácticas Específicas*, CEDE/UNSAM, arXiv:0809.0013
ROGERS J H, 1998, 'Origins of the ancient constellations', *Journal of the British Astronomical Association*, 108: 9-28.

Alejandro Gangui
Doctor en astrofísica, International School for Advanced Studies, Trieste, Italia.
Investigador adjunto del CONICET. Profesor de la FCEyN, UBA.
Miembro del Centro de Formación e Investigación en Enseñanza de las Ciencias, FCEyN, UBA.
*gangui@df.uba.ar*
*cms.iafe.uba.ar/gangui*

*Figuras*

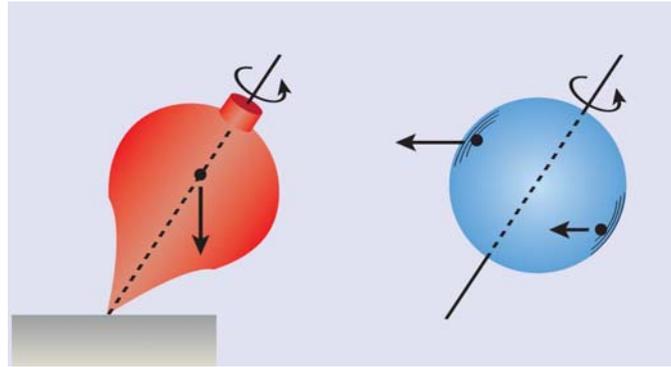

Figura 1. Comparación entre la fuerza que hace caer a un trompo en rotación y aquellas que gobiernan la inclinación del eje de rotación del globo terrestre. Una vez apartado el trompo de su posición perfectamente vertical, debido a perturbaciones diversas difíciles de evitar, comienza a caer hacia la tierra atraído por la fuerza gravitatoria de esta, es decir, por el propio peso del trompo (señalado con una flecha hacia abajo en la imagen de la izquierda). El caso de la Tierra es análogo, aunque esta no tiene hacia dónde caer, excepto hacia los demás cuerpos astronómicos que la rodean. La Luna, que es el cuerpo más cercano, ejerce fuerzas diferentes sobre distintas partes del abultamiento ecuatorial de nuestro planeta, como se muestra en la imagen de la derecha. El resultado es una competencia entre fuerzas con efectos opuestos aplicadas en distintos lugares de la Tierra. El efecto neto de esa competencia tiende a restituir al eje de la Tierra a una posición perpendicular al plano de la eclíptica. Nótese que estamos despreciando la pequeña diferencia de inclinación entre el plano de la eclíptica y el plano de revolución de la Luna alrededor de la Tierra.

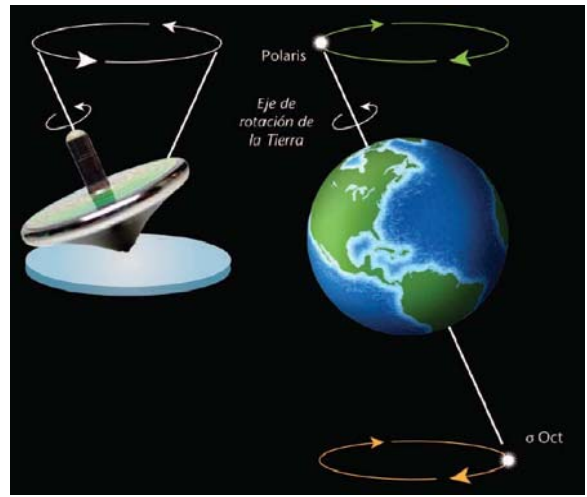

Figura 2. Movimientos de precesión de un trompo y del globo terrestre. Nótese la similitud entre ambos movimientos. Pero hay una diferencia. El trompo precesa en el mismo sentido que rota. La Tierra, en cambio, rota sobre su eje en un sentido que percibimos en su superficie como de oeste a este (por ello la frase "el Sol sale por el este"), pero su movimiento de precesión es en sentido contrario: de este a oeste. ¿Por qué esa diferencia? La respuesta viene dada por cómo operan las fuerzas ilustradas en la figura 1: el peso del trompo genera un torque que tiende a *alejar* su eje de rotación de la vertical; las fuerzas gravitatorias que ejerce la Luna sobre diferentes partes de la Tierra producen un torque que tiende a *acercar* el eje de rotación terrestre a la posición vertical. Torques contrarios generan movimientos de precesión contrarios.

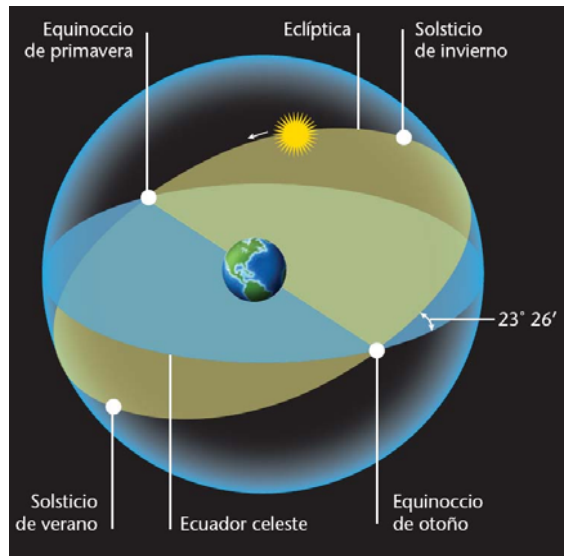

Figura 3. La bóveda celeste, el ecuador, la eclíptica, los solsticios y los equinoccios. El plano de la eclíptica viene definido por la trayectoria aparente del Sol en la bóveda celeste a lo largo del año. Forma un ángulo de unos 23º26' (23,5 grados aproximadamente) con el plano del ecuador celeste. Allí donde se cruzan esos planos imaginarios están los puntos equinocciales, separados 180º entre sí. A 90º de ellos sobre la eclíptica están los solsticios. Los solsticios y equinoccios señalados corresponden al hemisferio sur.

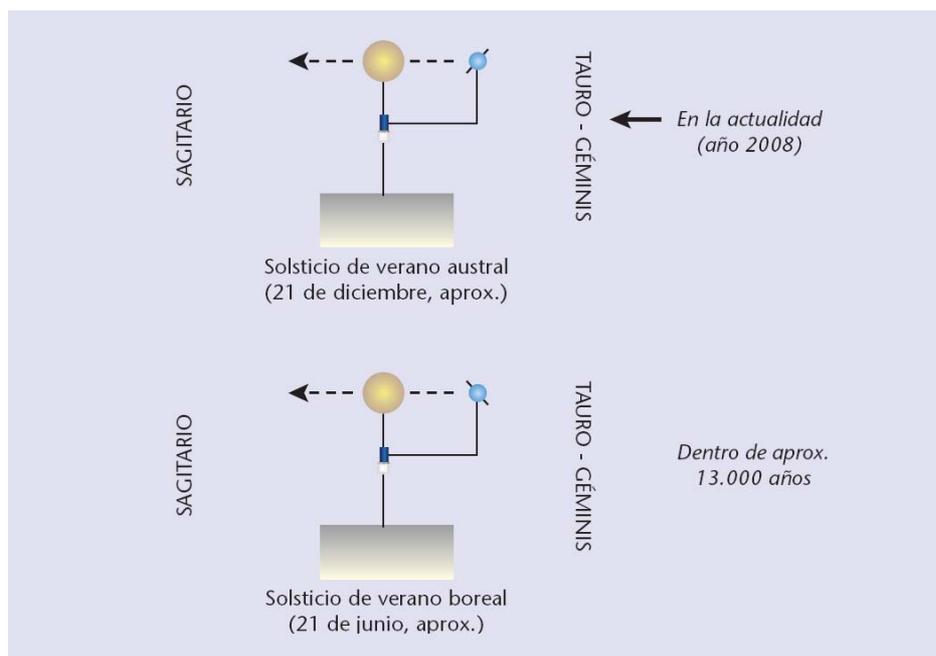

Figura 4. Con la maqueta de un sistema solar simplificado descripta en la nota didáctica del número 106 de CIENCIA HOY se pueden representar momentos singulares del año astronómico, como solsticios y equinoccios, ya sea en la época presente (croquis superior) como en cualquier otra época de la historia. En el dibujo inferior se representa la situación dentro de 13.000 años, para cuando el movimiento de precesión del eje de la Tierra habrá hecho que, al coincidir la trayectoria aparente del Sol con las estrellas de Sagitario, el calendario no registre el solsticio del verano austral (como en la imagen superior) sino el del verano boreal. Nótese que en la imagen inferior estamos suponiendo que la distribución de las estrellas del zodíaco, en particular las de la constelación Sagitario, no cambia demasiado en un lapso de más de diez mil años, algo que no es cierto pero no altera el ejercicio didáctico.

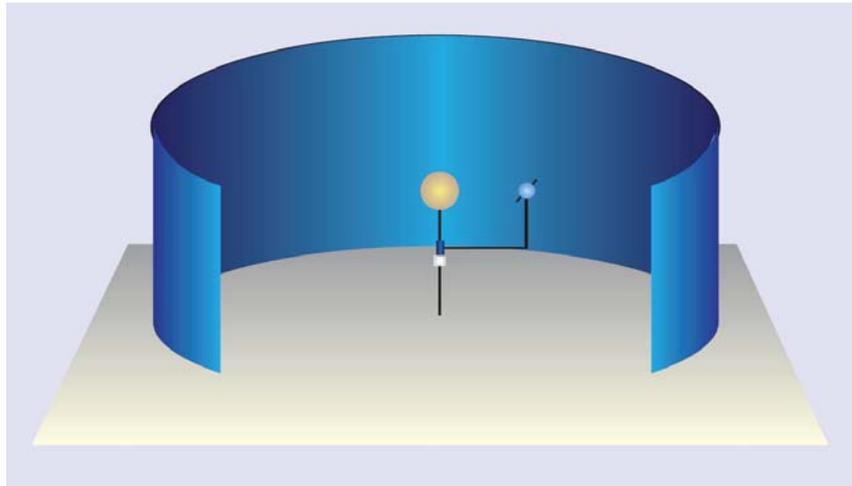

Figura 5. Maqueta del sistema solar simplificado con la banda zodiacal que lo rodea. Desde nuestro punto de observación en la Tierra, vemos el cielo estrellado como una inmensa esfera, dentro de la cual y, mas precisamente, en su centro, habitamos. Sobre ese telón de fondo de la bóveda celeste, las estrellas son puntos fijos (si prescindimos de la rotación de la Tierra), porque están tan lejos que no percibimos sus movimientos relativos a nosotros. El Sol, en cambio, parece desplazarse por un camino bien definido, al que llamamos la eclíptica. (En realidad, ese desplazamiento aparente del Sol es la consecuencia de que, al girar la Tierra en órbita a su alrededor, van cambiando las estrellas que veríamos −si su luz no nos lo impidiera− detrás de él). Con los planetas sucede algo más complicado, porque ellos también giran en órbita alrededor del Sol. Este, a lo largo de su camino aparente (por el que avanza al ritmo de casi un grado por día, pues completa 360º en poco más de 365 días), atraviesa grupos de estrellas o constelaciones que reciben desde antiguo el nombre de constelaciones zodiacales o zodíaco, una palabra de etimología incierta. Los nombres y símbolos de esas constelaciones son los signos del zodíaco. Nótese que la maqueta no respeta la escala del sistema real.

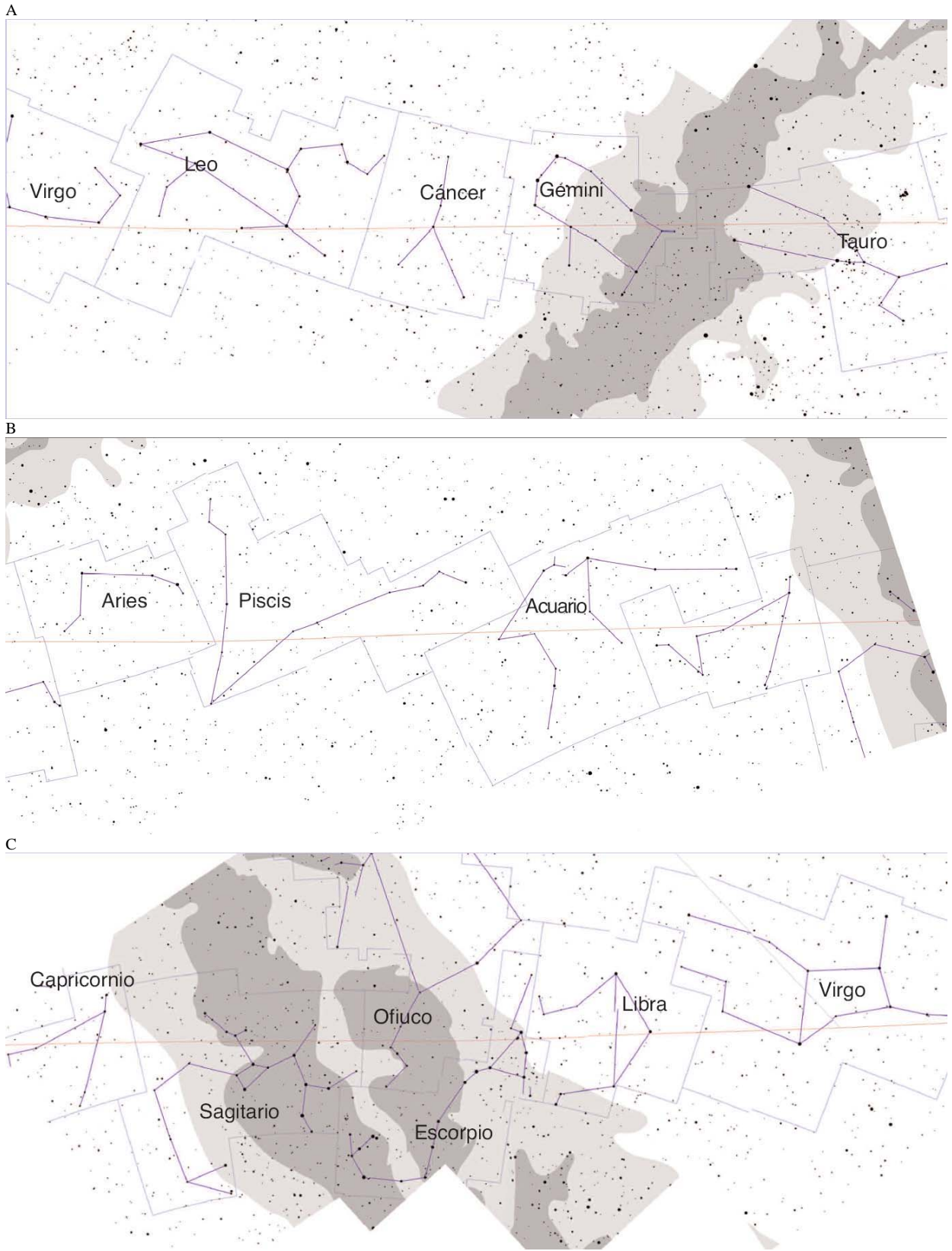

Figura 6. Tres tiras para fotocopiar y pegar con el propósito de constituir la banda zodiacal indicada en la figura 5. Identificadas con las letras A, B y C, se ensamblan una a continuación de la otra en ese orden. Adviértase que en la reproducción de las constelaciones hay un poco de superposición, para facilitar el ensamble. Se sugiere hacer una fotocopia ampliada de estas tiras (por lo menos un 20% más grandes que los originales, que se hallan aquí: http://cms.iafe.uba.ar/gangui/didaastro/ch/). Ello permitirá obtener una banda de mayor circunferencia y grosor, y facilitará su empleo como se indica en la figura 5.